\documentclass[12pt,english]{article}
\usepackage[T1]{fontenc}
\usepackage[latin1]{inputenc}
\usepackage{amsmath}
\usepackage{epsfig}
\usepackage{amsbsy}
\usepackage{rotating}
\usepackage{graphicx}
\usepackage[dvips]{color}
\usepackage{epsfig}
\makeatletter

\catcode`@=12
\usepackage{babel}
\makeatother
\begin{document}
\thispagestyle{empty}

\mbox{}
\vspace{0.75in}

\begin{center}
\textbf{\large Gravitational correction to SU(5) gauge coupling unification\\}
 \vspace{1.0in}
 \textbf{Jitesh R. Bhatt, Sudhanwa Patra and Utpal Sarkar}\\
 \vspace{0.2in} \textsl{ Physical Research Laboratory, Ahmedabad
380009, India}\\
 \vspace{.75in}
 \end{center}

\begin{abstract}

The gravitational corrections to the gauge coupling constants of
abelian and non-abelian gauge theories has been shown to diverge
quadratically. Since this result will have interesting consequences,
this has been analyzed by several authors from different approaches.
We propose to discuss this issue from a phenomenological approach.
We analyze the SU(5) gauge coupling unification
and argue that the gravitational corrections to gauge coupling
constants may not vanish when higher dimensional non-renormalizable
terms are included in the problem.

\end{abstract}

\newpage
\section{Introduction}
\hspace{1cm} The question of gravitational corrections to the
evolution of the gauge coupling constant has attracted some
attention in recent times, following the seminal paper of
Robinson and Wilczek \cite{cp1}. They studied the one-loop
quantum corrections to the running of the gauge couplings
in an effective quantum theory of gravity, which is valid
at energies below the Planck scale and found a quadratic
divergent behavior. The character of the correction has
been arrived at from a general consideration, which has been
shown to have important phenomenological consequences in
theories with low scale gravity \cite{gog}. However, this   
result has been questioned by some authors and the result
has been studied from different approaches. This gravitational
correction has been shown to depend on the choice of gauge
in an explicit calculation \cite{gaugedep}. They studied
the abelian theory and used a parameter dependent gauge to
arrive at their result. Subsequently a more general result
has been obtained using a gauge invariant background field
method that the gravitational corrections to the
gauge couplings vanishes \cite{toms}. Following the doubts
raised by these two references on the result of ref. \cite{cp1},
a one-loop diagrammatical calculation has been performed
in the full Einstein-Yang-Mills system, which had also confirmed
the vanishing of the one-loop contributions of quantum gravity to the
gauge coupling evolution \cite{cp2}.

The quantum gravity corrections to the running of gauge couplings
were calculated for pure Einstein-Yang-Mills system. Although our preliminary
results show that even after including scalar fields, the diagramatic
techniques would give vanishing gravitational correction, it is not
clear if the general results of ref. \cite{cp1} will be valid in some
cases. Recently the gravitational corrections to the gauge coupling
evolution has been studied including a cosmological constant and
quantum gravity effect has been found to affect the running of the
gauge couplings \cite{cc}.
However, the one-loop contributions in the presence of a cosmological
constant differs from that of ref. \cite{cp1}, which was obtained
from a general consideration.
This raises the question: what are the other factors that would make the
quantum gravity effects significant?

In this article we argue from a phenomenological approach that the
quantum gravity effects should be significant when higher dimensional
non-renormalizable interactions are taken into consideration. Since
quantizing the general theory of relativity for small fluctuations
around flat space gives us a non-renormalizable field theory, we need
to include an infinite set of higher dimensional counterterms. Since
these terms are suppressed by appropriate powers of the Planck mass
$M_p \sim 10^{19}$ GeV, at energies well below the Planck scale these
higher dimensional terms may be considered as small perturbations in
the effective theory of quantum gravity \cite{don}. However, at the
scale of grand unification these terms may not be ignored, and hence,
in some version of the grand unified theories dimension-5 and dimension-6
gauge invariant terms have been included on phenomenological ground to
see if these terms can change any of the conclusions for some reasonable
values of the coupling constants \cite{cp5}. It was found that although
the minimal SU(5) grand unified theory fails to satisfy the gauge
coupling unification, inclusion of the higher dimensional terms change
the boundary conditions and allow gauge coupling unification at a higher
scale \cite{cp5,cp3}.
Here we point out that if the gravitational contributions to the
gauge coupling evolution vanish, then the boundary conditions appearing
due to the higher dimensional terms become inconsistent. We then show
how the gauge coupling constants evolve from low energy
to the GUT scale and satisfy the non-renormalizable operator induced
matching condition at the new GUT scale, if we include gravitational
corrections to the gauge couplings, which diverge quadratically near
the Planck scale.

\section{Effect of higher dimensional operators in SU(5) unification}

Most of the grand unified theories (GUTs) with intermediate
symmetry breaking scales can satisfy the experimentally observed constraints on
proton lifetime ($\tau_{p}$) for the p $\rightarrow e^{+} \pi^{0}$ mode and
the electroweak mixing angle $\sin^{2}\theta_{w}$\\
$$\tau_{p} \geq 3 \times 10^{32}~{\rm yr},~~~~~ \sin^{2} \theta_{w} = 0.230 \pm 0.005 \,. $$
The minimal SU(5) and other GUTs with no intermediate symmetry breaking
scale and no new particles beyond the minimal representations are ruled out as
they predict significantly lower values. In other words, with the present
range for the $\sin^2 \theta_w$, if we evolve the three gauge coupling
constants from the electroweak scale to the grand unification scale, they do
not meet at a point, and hence, there is no unification. In an interesting
proposal it was pointed out that since the grand unification
occurs at a scale $ M_{U} \geq 10^{15} $ GeV), which is close
to the Planck scale,
it is natural to expect that there could be significant modification to the
GUT predictions by gravity-induced corrections \cite{cp5}. These corrections
may allow gauge coupling unification, make proton stable, give correct
neutrino masses and proper charged fermion mass relations at the
GUT scale, even for the minimal SU(5) GUT. In this article we include the
higher dimensional terms to study the gauge coupling unification and
infer that the evolution of the gauge coupling constants should be
modified by the gravitational corrections.

We start with the $SU(5)$ Lagrangian and then the breaking of
$SU(5)$  group into the Standard Model group $SU(3)_{C} \times
SU(2)_{L} \times U(1)_{Y}$ via the Higgs field $\phi$, which
transforms under the 24-dimensional adjoint representation of
$SU(5)$. We write down the Lagrangian as a combination of the usual
four dimensional terms plus the new higher dimensional terms which
has been induced by the non-renormalizable interactions of
perturbative quantum gravity. Since the couplings of these terms
are not known, we cannot make any predictions at this stage, so
we look for consistent solutions for a reasonable range of the
unknown parameters. The SU(5) gauge invariant Lagrangian, including
higher dimensional terms can be written as
\begin{equation} L = L_{0} + \Sigma_{n=1}  L^{(n)}\end{equation}
where\\
\begin{equation} L_{0} = -\frac{1}{2} {\rm Tr}(F_{\mu \nu}  F^{\mu \nu})\end{equation}
Where the sum is over the higher dimensional operators. For the
present we shall restrict ourselves to only
five- and six-dimensional operators, which are:
\begin{equation} L^{(1)} = -\frac{1}{2 }\frac{\eta^{(1)}}{M_{Pl}} {\rm Tr}(F_{\mu \nu} \phi F^{\mu \nu})\end{equation}
\begin{eqnarray} L^{(2)} &=& -\frac{1}{2 }\frac{1}{M_{Pl}^{2}} \left[ \eta_{a}^{(2)}
{{\rm Tr}(F_{\mu \nu} \phi^{2} F^{\mu \nu})+{\rm Tr}(F_{\mu \nu} \phi F^{\mu \nu} \phi)}
\phantom{a \over b} \right. \nonumber \\ && \left.
+\eta_{b}^{(2)}
{\rm Tr}(\phi^{2}) {\rm Tr}(F_{\mu \nu}  F^{\mu \nu}) +\eta_{c}^{(3)} {\rm Tr}(F_{\mu \nu} \phi) {\rm Tr}(F^{\mu \nu} \phi) \right]
\end{eqnarray}
where \\
\begin{equation}
F^{\mu \nu} = \partial_{\mu} A_{\nu} -\partial_{\nu} A_{\mu}
-i g \left[A_{\mu},A_{\nu} \right]
\end{equation}
\begin{equation}
{(A_{\mu})^{a}}_{b} =A_{\mu}^{i} \left[ \frac{\lambda_{i}}{2} \right]^{a}_{b}
\end{equation}
and
\begin{equation}
{\rm Tr}\left( \lambda_{i} \lambda_{j} \right) = \frac{1}{2} \delta_{i j}\,.
\end{equation}
Here $A^{i}$ is the ith component of the gauge field,
$\lambda_{i} $ is the corresponding generator and $\eta^{n}$, n=1,2,... are the unknown parameters,
induced by gravitational corrections.

When the scalar $\phi$ acquires a vacuum expectation value ($vev$) and breaks the SU(5)
symmetry at the GUT scale, we may replace these fields in the above expressions by
its $vev$. This will give us the effective low energy theory with only dimension-4
interactions, but the effective gauge fields will be modified below the GUT scale.
We may define the new physical gauge fields below the unification scale to be
\begin{equation} A'_{i} = A_{i} (1 + \varepsilon_{i})^{1/2}\end{equation} and the modified coupling
constants including the higher dimensional operators as
\begin{equation} {\widetilde g_{3}}^{2}(M_{U}) = {g_{3}}^{2}(M_{U}) (1 + \varepsilon_{C})^{-1}\end{equation}
\begin{equation} {\widetilde g_{3}}^{2}(M_{U}) = {g_{2}}^{2}(M_{U}) (1 + \varepsilon_{L})^{-1}\end{equation}
\begin{equation} {\widetilde g_{1}}^{2}(M_{U}) = {g_{1}}^{2}(M_{U}) (1 + \varepsilon_{Y})^{-1}\end{equation}
The $g_{i}$ are the couplings in the absence of higher dimensional operators,
whereas $\widetilde g_{i}$ are the physical couplings which evolve down to the lower scales.
The value of the $ \varepsilon^{n}$ associated with the given operator of dimension n+4
may be expressed in the following way
\begin{equation}
\varepsilon^{n} = \left[ \frac{1}{\sqrt{15}} \frac{\phi_{0}}{M_{Pl}}\right] ^{n} \eta^{(n)}
\end{equation}
The vev $ \phi_{0}$ is related to $M_{U}$
\begin{equation} \phi_{0} = \left[ \frac{6}{5 \pi \alpha_{G}} \right]^{1/2}  M_U\end{equation}
The change in the coupling constants are then related to the $ \varepsilon^{n}$s through
the following equations
\begin{equation}
\varepsilon_{C} = \varepsilon^{(1)} + {\varepsilon_{a}}^{(2)}
+ \frac{15}{2} {\varepsilon_{b}}^{(2)}+....
\end{equation}

\begin{equation}
\varepsilon_{L} = -\frac{3}{2} \varepsilon^{(1)} + \frac{9}{4}
{\varepsilon_{a}}^{(2)} + \frac{15}{2} {\varepsilon_{b}}^{(2)}+....
\end{equation}

\begin{equation}
\varepsilon_{Y} = -\frac{1}{2} \varepsilon^{(1)} + \frac{7}{4}
{\varepsilon_{a}}^{(2)} + \frac{15}{4} {\varepsilon_{b}}^{(2)}+ \frac{7}{8} {\varepsilon_{c}}^{(2)}+....
\end{equation}
This shows how the effect of higher dimensional
operator modify the gauge coupling constants. The
Unification scale, $M_{U}$, is now defined through the new boundary
condition 
\begin{equation} {g_{3}}^{2} (1 + \varepsilon_{C}) = {g_{2}}^{2}
(1 + \varepsilon_{L})={g_{1}}^{2} (1 + \varepsilon_{Y})={g_{0}}^{2}\,.
\end{equation}
With this in mind, one may use the standard one loop
renormalization group (RG) equations
\begin{equation} {\alpha_{i}}^{-1}(M_{z})= {\alpha_{i}}^{-1}(M_{U})
+ \frac{b_{i}}{2 \pi} \log\left( \frac{M_{U}}{M_{z}}\right) \end{equation}
with the beta functions
$b_{1}=\frac{41}{10}$,$b_{2}=\frac{-19}{6} $,$b_{3}=-7 $.
We have taken $N_{f}$=3 and $N_{Higgs}$=1.

Solving the RG equations without any higher dimensional contributions
yield
\begin{equation}
\log\left( \frac{M_{U}}{M_{z}}\right)= \frac{6}{67 \alpha} \frac{1}{D} \left[
{1-\frac{8}{3} \frac{\alpha}{\alpha_{s}}}+{\varepsilon_{C}- \frac{5 \varepsilon_{Y}+3
\varepsilon_{L}}{3} \frac{\alpha}{\alpha_{s}}}\right]
\end{equation}

\begin{equation}\sin^{2}{ \theta_{w}} = \frac{1}{D}\left[{\sin^{2} {\theta_{w}}} ^{(5)}
- \frac{19}{134}     \varepsilon_{C} + \frac{1}{67}  \left(  21
+ \frac{41}{2} \frac{\alpha}{\alpha_{s}} \right)  \varepsilon_{L}
+\frac{95}{402} \frac{\alpha}{\alpha_{s}} \varepsilon_{Y}  \right]
\end{equation}

\begin{equation}
\frac{1}{\alpha_{G}} = \frac{3}{67} \frac{1}{D}
\left[ \frac{11}{3 \alpha_{s}} +\frac{7}{\alpha} \right]
\end{equation}

\begin{equation}
D = 1 + \frac{1}{67} (11 \varepsilon_{C} + 21 \varepsilon_{L} + 35 \varepsilon_{Y})
\end{equation}
Where the $ {\sin^{2} {\theta_{w}} }^{(5)} $ is the usual minimal SU(5) prediction
\begin{equation}
{ \sin^{2} {\theta_{w}} } ^{(5)} = \frac{23}{134} + \frac{109}{201} \frac{\alpha}{\alpha_{s}}
\end{equation}
In this case of minimal SU(5), the gauge coupling constants do not meet
at a point, and hence, unification is not possible. We now
show how this result gets modified by including higher dimensional terms.

We first consider only the following SU(5) invariant
non-renormalizable (NR) (dimension five) interaction term
\begin{equation} L_{NR} = -\frac{1}{2 } \left( \frac{\eta}{M_{Pl}}\right)
{\rm Tr}(F_{\mu \nu} \phi F^{\mu \nu}) \,,
\end{equation}
where $ \phi_{24}$ is the Higgs 24-plet, $ \eta$ is a dimensionless parameter
and $M_{Pl}$ is the Planck mass. Suppose the Higgs field acquires a vacuum expectation value(vev)
\begin{equation}
\left\langle \phi \right\rangle = \frac{1}{\sqrt{15}} \phi_{0} {\rm diag}[1,1,1,-\frac{3}{2},-\frac{3}{2}]
\end{equation}
The SU(5) gauge symmetry breaks to $SU(3)_{C} \times SU(2)_{L}
\times U(1)_{Y}$  at this scale because of non-invariance of the
Higgs field under the SU(5) symmetry. The presence of
non-renormalizable couplings modifies the usual kinetic energy
terms of the $SU(3)_c$, $SU(2)_L$ and  $ U(1)_Y$ gauge boson part of
the low-energy Lagrangian. The modified Lagrangian becomes
\begin{equation}
-\frac{1}{2} (1 + \varepsilon) {\rm Tr}({F_{ \mu \nu}}^{(3)}  {F^{\mu \nu}}^{(3)})
-\frac{1}{2}  (1 -\frac{3}{2} \varepsilon) {\rm Tr}({F_{ \mu \nu}}^{(2)}
{F^{\mu \nu}}^{(2)}) -\frac{1}{2}  (1 -\frac{1}{2} \varepsilon)
{\rm Tr}({F_{ \mu \nu}}^{(1)}  {F^{\mu \nu}}^{(1)})\,,
\end{equation}
where the
superscripts 3,2 and 1 refer to gauge field strengths of $SU(3)$, $
SU(2)$  and  $ U(1)$ respectively and $ \varepsilon$ is defined as
\begin{equation}
\varepsilon = \left[ \frac{1}{\sqrt{15}} \frac{\phi_{0}}{M_{Pl}}\right]  \eta\,.
\end{equation}
We used $\varepsilon^{(2)}=\varepsilon^{(3)}=0$ and
$ {\varepsilon^{(1)}} = \varepsilon = \eta {\phi}_{0}/( \sqrt{15} M_{U} ) $, so that
$\varepsilon_{C} = \varepsilon $, $\varepsilon_{L} = -\frac{3}{2}
\varepsilon $, $\varepsilon_{Y} = -\frac{1}{2} \varepsilon $.
Now, using these expressions, we get
\begin{equation}\frac{1}{\alpha_{G}} = \frac{11 {\alpha_{s}}^{-1} +21 \alpha^{-1}} {67-38 \varepsilon}\,,
\end{equation}
\begin{equation}
\log\left( \frac{M_{U}}{M_{z}}\right)= \frac{6 \pi} {67-38 \varepsilon}
\left[ \alpha^{-1} -\frac{8}{3}{\alpha_{s}}^{-1}  +\left( \frac{7}{3}
{\alpha_{s}}^{-1} +\alpha^{-1} \right) \varepsilon   \right]
\end{equation}
\begin{equation}
\sin^{2}{ \theta_{w}} =  \frac{1}{67-38 \varepsilon}
\left[ \frac{23}{2} +\frac{109}{3} \frac{\alpha}{\alpha_{s}} -
\left( 41 + \frac{116}{3} \frac{\alpha}{\alpha_{s}} \right)
\varepsilon \right]
\end{equation}
Taking the experimental values of $\alpha_{s} = 0.1088$, $\alpha = 1/127.54$,
it is possible to obtain a consistent choice  of
the parameters $\varepsilon_{C}$, $\varepsilon_{L}$,
$\varepsilon_{Y}$ which satisfy the constraints on $ \sin^{2}
{\theta_{w}} $ and $ M_{U}$. But the unification scale remains low and
the proton lifetime becomes less than the present experimental bound.
For central value of $ \sin^{2} {\theta_{w}}
(=0.2333) $, we obtain $\varepsilon^{(1)}= -0.0441$ and $ M_{U}=
3.8 \times 10^{13}$ GeV and the corresponding value of
$\alpha_{G}=0.0245 $. The lifetime of proton ($m_{p}$ is the
mass of the proton)
\begin{equation}
\tau_{p} = \frac{1}{{\alpha_{G}} ^{2}}
\frac{{M_{U}}^{4}}{{m_{p}}^{5}}
\end{equation}
then becomes too low to be consistent
with  experimental limits on $\tau_{p}$ for the given value of $ M_{U}$.
Hence, it is not possible to obtain a consistent solution with the
five Dimensional operator.

\begin{table}[h]\centering
\caption{Unification in SU(5) using gravity corrections}
\begin{tabular}{|c|c|c|c|}
\hline
$\epsilon_{C}$ &  $\epsilon_{L}$  &  $\epsilon_{Y}$    &   $M_{U}$   \\
\hline
0.04 & 0.0675  &  0.24  &  $10^{17}$ GeV  \\
\hline
0.3894 & 0.44  & 0.98  &  $10^{18}$ GeV        \\
\hline
1.3894 & 1.445 & 1.98  &   $10^{18.6}$ GeV       \\
\hline
\end{tabular}
\label{tab1}
\end{table}

If we now include both five and six dimensional terms, then there
are whole range of parameters that are consistent with the values
of $ \sin^{2} {\theta_{w}} $, $M_{U}$ and proton lifetime. We
present a few representative set of values that are consistent
with proton lifetime in table \ref{tab1}. So, from now on we shall
consider both dimension five and dimension six non-renormalizable
terms for our discussion.

\section{Evolution of gauge couplings including gravitational contributions}

In the last section we discussed the effect of higher dimensional
non-renormalizable interaction on the boundary condition, satisfied
by the gauge couplings. In fact, the effective gauge couplings get
modified at the time of GUT phase transition, which allows the
gauge coupling unification for some parameter range. If we now
start evolving the gauge coupling constants from low energy, when
the effects due to the higher dimensional terms are negligible,
we should be able to reach the new modified boundary condition
continuously. In other words, the modified
effective gauge couplings should evolve with energy in such a
way that at low energy they become the usual gauge couplings. If
we now assume that the gravitational corrections to the evolution of
the gauge couplings vanishes, then this transition is not possible.
On the other hand, if we consider that the gravitational corrections
are of the quadratic nature, as recommended in ref. \cite{cp1},
then it is possible to continuously evolve the gauge coupling
constant from the modified effective coupling near the GUT phase
transition scale to the low energy experimentally observed couplings.

In this section we shall first argue how the non-renormalizable
interactions could change the gravitational corrections to the
gauge couplings. Then
we shall demonstrate how the gauge coupling constants evolve from
low energy to the unification scale in the presence of the higher
dimensional contributions. Although the modified boundary
condition and its effect was studied by many authors, the running
of the gauge couplings from low energy to the unification scale
could not be studied. This is because the running of the gauge
couplings in the presence of gravitational corrections were not
considered.

As the gauge boson vertex has strength $g$ and gravity couple to
energy momentum with a dimensional coupling $\propto
\frac{1}{M_{Pl}}$, dimensional analysis implies that the running
of couplings in four dimensions will be governed by a
Callan-Symanzik $\beta $ function of the form
\begin{equation}
\beta(g,E) = \frac{d g}{d ln E} = -\frac{b_{0}}{{4 \pi}^{2}} g^{3}
+ a_{0} \frac{E^{2}}{{M_{Pl}}^{2}} g
\label{gg}
\end{equation}
where the first term is the non-gravitational contribution and the
2nd term is the gravitational contribution, as suggested in ref. \cite{cp1}.
This quadratic gravitational correction was then revisited in ref.
\cite{gaugedep,toms,cp2} and it was shown that this contribution vanishes.
We shall now argue that in the presence of non-renormalizable interactions,
this contribution may not vanish.

Following equations 8-11, we write down the effective coupling constant
at the GUT scale as
\begin{equation}
\widetilde g^{-2} = g^{-2} + C,
\end{equation}
where $C$ is the contribution coming from the non-renormalizable
interactions. We shall now argue that although the gauge coupling
evolution may not be affected by gravitational corrections (as
stated in refs. \cite{gaugedep,toms,cp2}), the
evolution of $C$ is dominated by gravitational correction, and hence,
it should evolve as suggested in ref. \cite{cp1}.

\begin{figure}[!h]
 \centering
 \epsfig{file=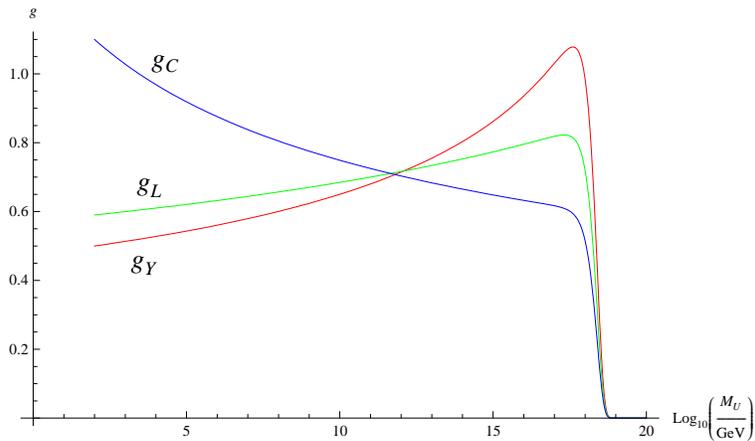, width=0.6\textwidth}
 \caption{Evolution of the gauge coupling constants without higher dimensional
 terms, but including gravitational corrections \cite{cp1}.}
 \label{fig2}
\end{figure}

In the absence of non-renormalizable interactions and gravitational
corrections, the three gauge couplings for a particular
model evolve as inverse logarithm of E at one loop order. Although
unification may not be achieved in case of minimal SU(5), including non-renormalizable terms
(i.e., including $C$) they may get unified 
at a scale  $\approx 10^{17-18}$ GeV.
In ref. \cite{cp1}, it was shown that in absence of $C$,
the couplings are unified near the Planck
scale and the value of the couplings are zero, as shown in
figure \ref{fig2}.
The negative value of $a_{0}$ in the beta function signifies that the
gravitational correction works in the  direction of asymptotic freedom,
i.e. it causes coupling constants to decrease at high energy (above $10^{16}$ GeV).

The modifications to the gauge couplings
arising due to non-renormalizable terms are symbolically
denoted by $C$ in equation 33.
To comply with
the unification condition described by equation 17,
the correction of each of the three coupling constants
will have different weights. This would give
nonzero contribution to the coupling constants unlike
in ref. \cite{gaugedep, toms,cp2}.
One can justify this point as follows:
For the purpose of a demonstration consider the diagramatic
method of ref. \cite{cp2}. Here one starts with the 
Einstein-Yang-Mills Lagrangian
\begin{equation}
{\cal L}_4 = {2 \over \kappa^2} \sqrt{-g} {\mathbf R} - {1 \over 2}
\sqrt{-g} g^{\mu \rho} g^{\nu \sigma} {\rm Tr} [F_{\mu \nu}
F_{\rho \sigma}] \,,
\end{equation}
with the Ricci scalar ${\mathbf R}$. We then expand the metric
in terms of the flat metric $\eta_{\mu \nu}$ and the graviton
field $h_{\mu \nu}$ to write
\begin{eqnarray}
g_{\mu \nu} &=& \eta_{\mu \nu} - \kappa h_{\mu \nu} + \kappa^2
h_{\mu \beta} h_{\nu}^{\beta} \nonumber \\
\sqrt{-g} &=& 1 + {\kappa \over 2} h + {\kappa \over 8} \left(
h^2 - 2 h^{\alpha \beta} h_{\alpha \beta} \right) \,.
\end{eqnarray}
It is then possible to write down the propagators for this
theory and explicitly calculate the one-loop diagrams to show
that the gravitational corrections to the $\beta$-functions
vanish \cite{gaugedep,toms,cp2}.
It should be noted that the  term of 
type $\sqrt{-g} g^{\mu \rho} g^{\nu \sigma} {\rm Tr} [F_{\mu \nu}
F_{\rho \sigma}]$  
 (in equation 34)  give contribution to the coupling
constant that is  quadratic in the energy \cite{cp1}.

If we now include the scalar fields $\Phi$ in the theory, there
will be interactions of the scalar fields with the graviton
field, which comes from the Lagrangian
\begin{equation}
{\cal L}_S = \sqrt{-g} [D_\mu \Phi D_\nu \Phi] g^{\mu \nu} \,.
\end{equation}
In this case also there seem to be cancellation of the quadratic
divergences (we considered the diagrams to order $\kappa^2$ for
the abelian case only) and there may not be any gravitational
corrections to the gauge coupling evolution.

 However, the inclusion of  higher
dimemsional non-renormalizable terms would completely
change the scenario.
Such non-renormalizable terms are expected in a
theory that incorporates the effect of quantum gravity.
In any grand unified theory, where the unification scale is only
2-3 orders of magnitude lower than the Planck scale (the
proliferation of particles near the GUT scale could also
lower the Planck scale \cite{cx}), such
non-renormalizable terms may contribute significantly.
Consider, for example, the dimension-5
term in presence of the 24-plet scalar $\phi$ of SU(5)
\begin{equation}
{\cal L}_5 = - {1 \over 2 M_{Pl}}
\sqrt{-g} g^{\mu \rho} g^{\nu \sigma} {\rm Tr} [F_{\mu \nu}
F_{\rho \sigma} \phi] \,.
\end{equation}
For the case when $E\leq {M}_U$,
the scalar $\phi$ acquires a $vev$ ($\langle \phi \rangle
\equiv M ~{\rm diag}[1,1,1,-3/2, -3/2]$), 
\newpage
this term would give contribution to the $C$ term in equation 33 
that vary quadratically with the energy.
However, to be consistent with the modified boundary condition
given by equation 17, the different gauge
fields with different weight factors 
will give nonzero contribution. It ought to be noted
that the coupling constants now meet at $E\approx M_U$ 
which is lower than the Planck scale 
This supports our earlier inference that the
gravitational corrections to the gauge couplings may not vanish
when the higher dimensional interactions are included.
Above the unification scale $ {{M_U}}$, the scalar field has not
acquired $vev$ and SU(5) symmetery is exact.
In this regime there will be only one gauge coupling
constant for entire SU(5) and it will evolve without
any gravitational corrections as if
the higher dimensional terms were absent.

Figure (2) shows how the coupling constants vary
with energy in the presence of $C$ terms in the regime
$E\leq M_U$. 
For the regime $E\ge {{M_U}}$, there is only one coupling
constant as the exact SU(5) symmetry is restored. 
In this case there will be
no gravitational corrections in this case
as pointed out in ref. \cite{gaugedep,toms,cp2}.

\begin{figure}[h]
 \centering
 \epsfig{file=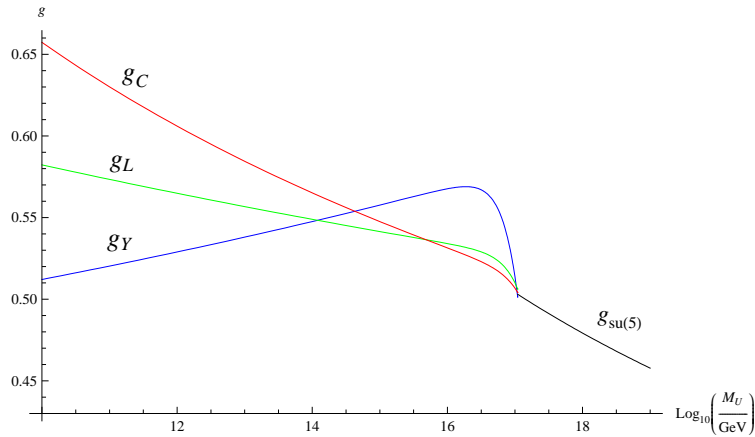, width=0.6\textwidth}
 \caption{Evolution of the gauge coupling constants in the presence of
 higher dimensional terms and gravitational corrections.}
 \label{fig3}
\end{figure}

\section{Conclusion}

The higher dimensional effective contributions has been studied in
the literature, whereby the gauge coupling constants get modified
near the grand unification scale. These modifications of the
boundary conditions allow gauge coupling unification even for the
minimal SU(5) GUT. However, the running of the modified gauge
couplings have not been studied. We show that this modified gauge
couplings should evolve including the gravitational corrections,
otherwise the low energy gauge couplings may not be consistent with
the modified boundary conditions. From this we infer that
the gravitational corrections to the gauge couplings may not
vanish when higher dimensional non-renormalizable interactions
are included in the Einstein-Yang-Mills system.

\end{document}